\documentclass[showpacs,
prb,aps,eqsecnum,groupedaddress]{revtex4}
\usepackage{amsmath,amssymb,latexsym,amsfonts}
\usepackage[dvips]{graphicx}
\begin{document}
\title{Ballistic quantum transport at high energies and high magnetic fields}
\author{S.~Rotter}
\author{B.~Weingartner}
\author{N.~Rohringer}
\author{J.~Burgd\"orfer}
\affiliation{Institute for Theoretical Physics, 
Vienna University of Technology,
Wiedner-Hauptstr.~8-10, A-1040 Vienna, Austria} 
\date{\today}
\begin{abstract}               
We present an extension of the modular recursive Green's function method 
(MRGM) for ballistic quantum transport to include magnetic fields. 
Dividing the non-separable two-dimensional scattering problem 
into separable substructures allows us to calculate transport
coefficients and scattering wavefunctions very efficiently.  
Previously unattainable energy and magnetic field 
regions can thereby be covered with high accuracy. The method is applied
to  magnetotransport through a circle and a stadium shaped quantum dot 
at strong magnetic fields and high energies.  
In the edge state regime we observe strong multi-frequency Aharonov-Bohm 
oscillations. By analyzing them in terms of a multi-channel 
interference model, we classify these fluctuations within the 
framework of Fano resonances and discuss their geometry independence. 
For high energies (mode numbers) we observe localization 
of the scattering wavefunction near classical trajectories. 
\end{abstract}
\maketitle
\section{INTRODUCTION}
Accurate simulations of ballistic transport through quantum dots 
have remained a computational challenge, despite the conceptional 
simplicity of the problem. This is in part due to the fact that many 
of the most interesting phenomena occur in a parameter regime of
either high magnetic field $B$ or high Fermi energy $E_F$. 
The regime of strong magnetic field $B$, where the 
magnetic length (in a.u.) $l_B=\sqrt{c/B}$ is small compared 
to the linear dimension $D$ of 
the dot, $l_B \ll D$, gives rise to the Quantum Hall effect,\cite{pran87}  
the Hofstadter butterfly,\cite{hofs76} and Aharonov-Bohm 
oscillations of transport coefficients.\cite{ashc76} The 
high energy domain, where the De Broglie wavelength 
$\lambda_D=\sqrt{2E_F}$ 
satisfies $\lambda_{D} \ll  D$, is of particular relevance for approaching 
the semiclassical limit of quantum transport and for 
investigations of quantum signatures of classical 
chaos.\cite{gutz91,stoe99,hell84} Both of 
these regimes pose considerable difficulties for a numerical 
treatment. Methods based on the expansion of the scattering 
wavefunction in plane or spherical waves become 
invalid at high fields since diamagnetic contributions are 
generally neglected.\cite{yang95} Methods employing a 
discretization on a grid are limited by the constraint that 
the magnetic flux per unit cell must be small, which, in turn, 
requires high grid densities for large $B$.\cite{mack83}
The same requirement has to be met for high $E_F$,
where many grid points are needed to accurately describe the 
continuum limit. This implies however a large number of
inversions of high-dimensional matrices and therefore limits
the applicability of these approaches for large $B$ and (or) 
large $E_F$.\\In the current communication we propose 
an approach that allows accurate treatment 
of these regimes. We present an extension of the previously\cite{rott00}  
introduced {\it Modular Recursive Green's function Method} (MRGM) 
to include an additional magnetic field perpendicular to the
two-dimensional scattering surface. The underlying 
idea for our approach goes back to Sols \textit{et al.}\cite{sols89}
and to the widely used
\textit{Recursive Green's Function Method} (RGM).\cite{mack83,ferr97}  
In the standard RGM the Green's
function is propagated through the scattering region from one transverse strip
to the next by repeated solutions of a matrix Dyson equation.
We show that the efficiency of this conventional discretization 
can be increased considerably
by taking the symmetries of a scattering problem into account from the outset.
Specifically, when the two-dimensional nonseparable open 
quantum dot can be built
up out of simpler separable substructures (referred to in the 
following as modules), the Green's functions for each of these 
modules  can be calculated efficiently and virtually exactly. 
Calculation of the $S$ matrix and of the scattering wavefunction 
is thus reduced to ``welding'' together the modules by a very 
small number of recursions. Key to this approach are 
tight-binding grids which are symmetry-adapted for each module. 
This leads to the separability of the eigenfunctions in the modules 
and allows an efficient incorporation of boundary conditions. 
As a result, a much higher 
grid density can be easily handled, which, in turn, is 
prerequisite for treating short magnetic lengths 
$l_B$ and short wave lengths $\lambda_D$. 
Matrix Dyson equations have to be solved only for  
each junction between the modules. The total number of 
necessary recursions  (i.e.~high-dimensional matrix inversions) 
is thereby reduced to the number of modules needed to 
reconstruct the quantum dot.\\The efficiency of the
MRGM will be demonstrated by applying it to transport through a circular
and a stadium shaped quantum dot. These systems are known as prototype 
structures for
regular and chaotic dynamics and have been studied thoroughly in the
literature.\cite{gutz91,stoe99,hell84,marc92}
Concerning the theoretical approaches for the investigation
of electron dynamics in quantum dots, considerable attention has been
dedicated to reach higher 
energies\cite{simo97,bies01,zozo96,akis97}
and higher magnetic fields.\cite{zozo96,lent91,ji95,chris98,horn02}
Especially for the study of transport through open stadium billiards,
several different methods have been 
employed.\cite{naka92,naka94,wang94,yang95,akis97,nazm02}
In the following we will present numerical results obtained by
the MRGM which attain a parameter range, to our knowledge
not yet explored by other approaches.
For small $\lambda_{D}$ we investigate the localization 
of the scattering wavefunction near classical scattering 
trajectories. Characteristic differences in the dynamics 
of generically regular and chaotic systems will be highlighted. 
In the high magnetic field regime, which is governed 
by edge states, differences between the dynamics in 
different geometries disappear 
and are replaced by universal quasi-periodic conductance oscillations. 
At a critical magnetic field these oscillations break off and transport
terminates entirely.
In the regime where more than one edge state is excited in the dot,
we find interference fluctuations 
which we analyze in terms of a multi-channel Fano interference  
model.\cite{kim01} The key to the understanding of the observed
fluctuations is that inter-channel scattering between 
different edge states takes place only by diffractive scattering  
at the lead junctions.\\This paper is organized as follows. 
In section II the method 
for inclusion of a magnetic field in the MRGM is presented. Section III
is dedicated to a discussion of numerical results, illustrating the 
high magnetic field and high energy behaviour in quantum dots. 
The paper concludes with a short summary
in section IV.
\section{METHOD}
We consider ballistic nanostructures with a
constant electrostatic potential inside the two-dimensional cavity,  
impose hard-wall
boundary conditions, and assume a constant magnetic field to be oriented
perpendicular to the scattering plane.
The shape of the quantum dot will be chosen to be either a circle
or a stadium (see Figs.~\ref{fig:highenergy},\ref{fig:edgepics} 
below), which represent prototype systems for 
regular and chaotic classical dynamics, respectively.
Two semi-infinite 
waveguides of width $d$ at different electrochemical potentials 
($\mu_1,\,\mu_2$) are attached. The aperture of the 
leads is chosen to be very small 
$d/D=d/\sqrt{A^{\rm dot}}=0.0935$, 
where $A^{\rm dot}=4+\pi$ is the scaled area of all the 
cavities studied and $D$ is a characteristic linear 
dimension of the cavity. At asymptotic distances, i.e.~far
away from the quantum dot, scattering boundary conditions 
are imposed. The asymptotic scattering state can be factorized 
into a longitudinal flux-carrying plane wave and a transverse 
standing wave. The latter is a simple sine wave in the field-free 
case and a combination of Kummer functions when 
the magnetic field is turned on.\cite{peet88,schu90}
In our local coordinate system
the longitudinal (transverse) direction in the $i$-th lead is
always denoted by $x_i$ $(y_i)$. The wavefunctions in the 
waveguides thus vanish at  $y_i=\pm d/2$.
Atomic units $(\hbar=|e|=m_{\rm eff}=1)$ 
will be used from now on, unless explicitly stated otherwise.
\subsection{Brief review of the MRGM}
In order to highlight  the technical difficulties in incorporating a
magnetic field we start by briefly reviewing the MRGM for the 
field-free case. Starting point is the observation that a 
large class of dot geometries with non-separable boundaries can 
be decomposed into separable two-dimensional substructures, 
referred to in the following as  modules. For each of these 
modules the discretization of the corresponding tight-binding 
(tb) Hamiltonian can be performed on a symmetry-adapted grid. 
The grid for each module is chosen such that the eigenfunctions 
of the tb Hamiltonian 
\begin{equation}\label{eq:tbhamilton}
  \hat{H}^{\rm tb}=\sum_i\varepsilon_i\,|\,i\,\rangle\langle\,i\,|
+\sum_{i,j}V_{i,j}\,|\,i\,\rangle\langle \,j\,|
\end{equation}
separate into two generalized coordinates. $\hat{H}^{\rm tb}$ 
contains hopping potentials $V_{i,j}$ for nearest-neighbour coupling 
and site energies $\varepsilon_i$. Both quantities are chosen such 
that the Schr\"odinger equation, 
$\hat{H}^{\rm tb}|\psi_m\rangle=E_m|\psi_m\rangle$, 
converges towards the continuum Schr\"odinger equation in the 
limit of high grid point density. The most 
straightforward application of this approach refers to modules 
for which the boundaries are nodal lines of 
Cartesian ($x,y$) or polar coordinates ($\varrho,\varphi$).
For these coordinate systems we have \cite{rott00} at $B=0$
\begin{eqnarray}
\begin{array}{rclrclrcl}
 V^x_{i,i\pm 1}&=&\frac{-1}{2\Delta x^2},\,&V^y_{j,j\pm 1}&=&\frac{-1}
{2\Delta y^2},\,&\varepsilon_i&=&\frac{1}{\Delta x^2}+\frac{1}{\Delta y^2},\\
V^\varrho_{i,i\pm 1}&=&\frac{-\varrho_{i\pm1/2}}{2\varrho_i\Delta\varrho^2},\,&
V^\varphi_{j,j\pm 1}&=&\frac{-1}{2\varrho_i^2\Delta\varphi^2},\,&
\varepsilon_i&=&\frac{1}
{\Delta\varrho^2}+\frac{1}{\varrho_i^2\Delta\varphi^2},
\end{array} 
\end{eqnarray}
with $\varrho_i=|i-1/2|\Delta\varrho$.
For separable energy eigenfunctions of the general form $| 
E_m\rangle=|E_k\rangle\otimes|E_{k,n}\rangle$ the spectral representation 
of the retarded $(+)$ and advanced $(-)$ Green's function 
$G^\pm({\bf r},{\bf r'},B,E_F)$ of the module is simply given by
\begin{equation} \label{2.3}
G^\pm({\bf r},{\bf r'},B,E_F)=
\sum_k\langle \alpha|E_k\rangle\langle E_k|\alpha'
\rangle\sum_n\frac{\langle \beta|E_{kn}\rangle\langle E_{kn}|\beta'\rangle}{E_F
\pm i\epsilon-E_{kn}}.
\end{equation}
where $(\alpha,\beta)$ stand for the (generalized) coordinates 
$(x,y)$ or $(\phi,\rho)$. The indices $(k,n)$ represent the quantum 
numbers of the separable eigenfunctions 
$|E_k\rangle,\,|E_{k,m}\rangle$ associated with 
the degrees of freedom $\alpha$ and $\beta$ respectively.\\The Green's 
functions of the separate modules are joined 
by solving a matrix Dyson equation,
\begin{equation}\label{2.4}
G=G^0+G^0\bar{V}G,
\end{equation}
where $G^0$ and $G$ denote Green's functions of the disconnected 
and the connected modules, respectively. The matrix $\bar{V}$ 
denotes the hopping potential $V$ multiplied  by the size of the unit
cell $\bar{V}=V \Delta_R$, which in a Cartesian (polar) grid is
$\Delta_R=\Delta x \Delta y\,\,(=\varrho_i \Delta \varrho \Delta 
\varphi)$. The complete scattering structure can thus be assembled 
from the individual modules (much like a jigsaw puzzle).
The number of necessary recursions [i.e.~solutions of (\ref{2.4})]  
is (approximately) equal to the number of modules. 
The exact number depends on the number of link modules required 
for different grid structures. For example, in order to connect 
a half-circle with a rectangle we need one additional link module 
which is plugged in between  [see 
Ref.~\onlinecite{rott00} for details].  
The key property of these link modules is their
adaption to two grid symmetries [see Fig.~\ref{fig:3}b]. 
Mathematically speaking, the transition
from a polar to a Cartesian grid requires  a link module in 
order to preserve the hermiticity of the tb Hamiltonian at the junction.
In the recursion the link module is connected to the Cartesian (polar)
grid by means of the hopping potential $\bar{V}^x(\bar{V}^\varphi)$, 
respectively.\\Once the Green's function $G^+$
for the combination of all modules is assembled,  the transmission 
amplitudes $t_{nm}$
from entrance lead mode $m$ into exit lead mode $n$ can be calculated by 
projecting $G^+$ onto the transverse wavefunctions 
in the leads $\chi_{n}(y_i)$. 
With the corresponding longitudinal wave numbers $k_{x_i,n}$ we have 
(at zero magnetic field), 
\begin{equation}
\label{2.5}
t_{nm}(E_F)=-i\sqrt{k_{x_2,n} k_{x_1,m}} \int^{d/2}_{-d/2} dy_2
\int^{d/2}_{-d/2} dy_1\ \chi_{n}^{*}(y_2)\,G^+(y_2,y_1,E_F)\,
\chi_{m}(y_1)\,.
\end{equation}
Together with the reflection amplitudes $r_{nm}$ 
(for which an analogous relation holds)
the $S$-matrix is completely determined and satisfies the
unitarity condition implied by current conservation,
\begin{equation}\label{eq:landauer1}
\sum_{n=1}^{M}(|t_{nm}|^2+|r_{nm}|^2)=1\,.
\end{equation}
The integer $M$ denotes the 
number of open channels in the leads.
According to the Landauer formula, the total conductance $g$ 
through the quantum 
dot is given by
\begin{equation}\label{eq:landauer2}
g=\frac{1}{\pi}\sum_{m,n=1}^{M}|t_{nm}|^2=
\frac{1}{\pi}\,T^{\rm tot}\quad{\rm with}\quad T^{\rm tot}+R^{\rm tot}=M\,.
\end{equation} 
\subsection{Inclusion of the magnetic field}
\begin{figure}[!tbh] 
\includegraphics[draft=false,keepaspectratio=true,clip,width=0.75\linewidth]{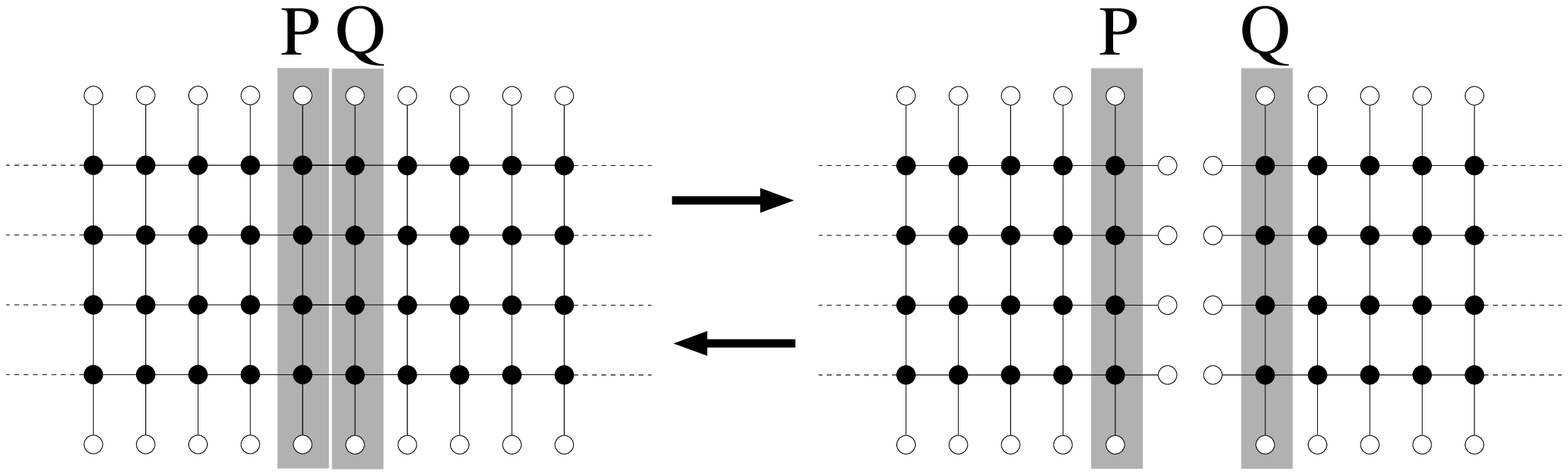}%
\caption{Joining and disconnecting of modules by application of 
a Dyson equation: two semi-infinite leads. The hard wall boundary 
conditions at the sites on the border of the modules are represented 
by empty circles (accessible space by full circles).
The gray shaded areas $P$ and
$Q$ are those grid slices at which the Green's functions 
are evaluated (see text).}
\label{fig:dyson1}
\end{figure}

Incorporation of the magnetic field into the MRGM poses a 
number of complications. The solutions of these difficulties will
be presented in this section. At the core of the problem is the preservation 
of separability of the Schr\"odinger equation. The usage of 
gauge transformations as well as of Dyson equations for 
decomposing non-separable structures into separable 
substructures plays a key role in accomplishing this goal. 
The field ${\bf B}=(0,0,B)$
enters   the tb Hamiltonian (\ref{eq:tbhamilton}) by means
of a Peierls phase factor,\cite{ferr97,peie33} 
\begin{equation}\label{IIA1}
 V_{\bf r,r'}\,\longrightarrow \, V_{\bf r,r'}
\cdot \exp \left[i/c \int_{\bf r}^{\bf r'}{\bf{A}(\bf{r})dr}\right]\,,
\end{equation}
with which  the field-free hopping potential $ V_{\bf r,r'}$ 
is multiplied. The   vector potential $\bf{A}(\bf{r})$ satisfies
$\nabla \times \bf{A} (\bf{r})=\bf{B}$. The Peierls phase
will, of course, in most cases destroy the separability
of the eigenfunctions of $\hat{H}^{\rm tb}$.
The difficulties can be, in part, circumvented by exploiting the 
gauge freedom of the vector potential, i.e., 
\begin{equation}\label{2.9}
{\bf A} \rightarrow {\bf A'} = {\bf A} + {\nabla} \lambda\,,
\end{equation}
where $\lambda ({\bf r})$ is a scalar function. By an appropriate 
choice of $\lambda $ the wavefunction remains separable on a 
given symmetry adapted grid. Specifically, to preserve separability
we employ the Landau gauge for a Cartesian grid 
\begin{subequations}
\begin{equation}
\label{2.10a}
{\bf A} = (-By, 0,0)\; ,
\end{equation}
and the ``symmetric'' or circular gauge for a polar grid
\begin{equation}
\label{2.10b}
{\bf A} = B/2(-y,x,0)={\mathbf \rho}\times{\bf B}/2\,.
\end{equation}
\end{subequations}
The scalar gauge potential generating the 
gauge transformation from (\ref{2.10a}) to (\ref{2.10b}) is 
$\lambda (x,y)=Bxy/2$.\\A major complication results from the 
fact, that in the presence of the magnetic field the separability 
on an unrestricted grid of a given  symmetry does not imply the 
separability in the presence of boundary conditions of the same 
symmetry. We illustrate this problem with the help of one 
typical example, the \textit {semi-infinite} quantum wire with 
lead width $d$ (Fig.~\ref{fig:dyson1}). 
We impose hard-wall boundary conditions 
$\psi (x,y = \pm d/2)=0$  and consider first the \textit{infinite} 
quantum wire along the $x$ direction. Because of the Cartesian 
boundary conditions, the symmetry adapted gauge is the Landau 
gauge ${\bf A} = -By\hat{\bf x}$. Consider, for notational simplicity, 
the Schr\"odinger equation in the continuum limit, 
\begin{equation}\label{2.11}
H\phi(y,x)=\frac{1}{2} \left( {\bf p}+\frac{1}{c}{\bf A}\right)^2 
\phi (x,y)=\frac{1}{2} \left(-\frac{\partial^2}{\partial x^2} 
-\frac{\partial^2}{\partial y^2} - \frac{i2B}{c}y 
\frac{\partial}{\partial x}+\frac{B^2y^2}{c^2} \right)
\phi (x,y)=E_F \phi(x,y).
\end{equation}
Since the longitudinal momentum $p_x = -i\partial /\partial x$ 
commutes with $H$, the separability of the wavefunction persists 
in the presence of the magnetic field: $\phi (x,y) = f_k(x) \chi (y)$ 
with $f_k (x)= e^{ikx}$. If, however, one introduces an additional 
Cartesian boundary condition along the $y$-axis [i.e.~$\psi(x=0, y)= 0$ 
for a semi-infinite lead] the situation changes. In the 
absence of the magnetic field, $B=0$, the linear term in $p_x$ 
vanishes and thus the choice $f(x)=\sin (kx)$ [i.e.~a linear 
combination of $f_{\pm k} (x)$] satisfies the boundary condition 
and preserves the separability, even though $p_x$ is no longer
conserved in the semi-infinite lead. However, for $B\neq 0$ and the 
same boundary condition $\psi(x=0,y)=0$, the term linear in $B$ 
and $p_x$ destroys the separability. The wavefunction takes now 
the general form
\begin{equation}
\label{2.12}
\phi(x, y)=\sum_{m} e^{ik_mx} \sum_{n} c_{mn} \chi_{mn}(y).
\end{equation} 
The breakdown of separability by the introduction of an additional 
boundary condition indicates that the Green's function of confined 
modules will be more complex than for extended systems for the 
same symmetry adapted grid and the same gauge. Therefore, the 
program of the modular method of building-up extended complex 
structures by ``welding together'' smaller modules of higher 
symmetry will be executed in reverse: non-separable confined 
modules will be generated by ``cutting in pieces'' larger 
separable modules.  Confining boundary conditions will be 
introduced rather than removed by the matrix Dyson equation. 
In the example above, the \textit{semi-infinite} quantum wire is generated 
by cutting the \textit{infinite} wire at the line $x=0$, thereby 
imposing the additional boundary condition. Just as connecting modules, 
so is disconnecting a given module equivalent to the application 
of a matrix Dyson equation, 
\begin{equation}
\label{2.13}
G^E=G^C+G^C \bar{V} G^E.
\end{equation}
In this context $G^E$ ($G^C$) is the Green's function of the 
extended (confined) module and $\bar{V}$ 
is the hopping potential that connects the modules. Solving 
(\ref{2.13}) \textit{in reversed mode} 
(i.e.~for $G^C$ rather than for $G^E$) 
amounts to dissecting the larger module.\\Provided 
that the Green's functions of all the necessary modules 
are available, we have to link them with each other to assemble 
the entire scattering geometry. However, in the presence of a 
magnetic field we have to take into account that the different
 modules will be calculated in different symmetry-adapted 
gauges. Joining different modules requires, therefore, in 
general a gauge transformation. For the Green's function on 
the grid $G({\bf r}_i, {\bf r}'_j)$ this transformation is 
simplified by the fact that the matrix of gauge transformations
\begin{equation}
\label{2.14}
\left[ \Lambda({\bf r}_j) \right]_{jk}=\exp 
\left[-i\lambda({\bf r}_j)/c \right] \delta_{jk}
\end{equation}
is diagonal in the grid representation. Correspondingly, the 
transformation of both the hopping potential $\bar{V}$ and 
the Green's function is local, i.e.
\begin{equation}
\label{2.15}
\bar{V}({\bf r}_i,{\bf r}'_j)\rightarrow \bar{V}'
({\bf r}_i,{\bf r}'_j)=\Lambda({\bf r}_i)\,
\bar{V}({\bf r}_i,{\bf r}'_j)\Lambda^*({\bf r}'_j)
\end{equation}
\begin{equation}
\nonumber
G({\bf r}_i,{\bf r}'_j)\rightarrow G'({\bf r}_i,{\bf r}'_j)=
\Lambda({\bf r}_i)\,G({\bf r}_i,{\bf r}'_j)
\Lambda^*({\bf r}'_j)\,.
\end{equation}
It is thus not necessary to transform   gauges of different 
modules to one global gauge. Instead, it is  sufficient to 
perform a local gauge transformation at the points of the 
junctions $\{{\bf r}_i\}$, such that the gauges of the two 
modules to be joined agree \textit{at these points}.\\Finally, 
in order to extract the $S$-matrix, i.e.~the amplitudes 
$t_{nm}$ and $r_{nm}$, matrix elements of the current operator 
must be of gauge-invariant form. This requirement
can be fulfilled by employing a double-sided 
gradient operator which is defined as\cite{bara89}
\begin{equation}\label{IIB6}
f\mathord{\buildrel{\lower3pt\hbox{$\scriptscriptstyle
\leftrightarrow$}}\over {\bf D}}g=f({\bf x}){\bf D}g({\bf x})-g({\bf x})
{\bf D}^*f({\bf x})=-g\mathord{\buildrel{\lower3pt\hbox{$\scriptscriptstyle
\leftrightarrow$}}\over {\bf D}}f\quad{\rm with }\quad {\bf D}=
{\nabla}-\frac{i}{c}{\bf A}({\bf x})\,.
\end{equation}
With its help the transmission amplitudes can be evaluated 
as\cite{bara89,ando91,zozo96}
\begin{eqnarray}\label{2.17}
t_{nm}(E_F,B)&=&-\frac{i}{4\sqrt{\theta_n\theta_m}}\,\int^{d/2}_{-d/2} dy_2
\int^{d/2}_{-d/2} dy'_1 \nonumber\,\chi_{n}^{*}(y_2)e^{-ik_{n}x_2}\,
(\mathord{\buildrel{\lower3pt\hbox{$\scriptscriptstyle
\leftrightarrow$}}\over {\bf D}}{\phantom{.}\!}\cdot\hat{\mathbf x}_2)^*\\
&&G^+({\mathbf x}_2,{\mathbf x}'_1,E_F,B)\,
(\mathord{\buildrel{\lower3pt\hbox{$\scriptscriptstyle
\leftrightarrow$}}\over {\bf D}}{\phantom{.}\!'}\cdot\hat{\mathbf x}'_1)
\chi_{m}(y'_1)e^{ik_{m}x'_1}\; .
\end{eqnarray}
The unit vectors $\hat{\mathbf x}_n$ are assumed to be pointing in 
outward direction of the $n$-th lead
and $\theta_m$ denotes the outgoing particle
flux carried by $\chi_{m}(y'_1)e^{ik_{m}x'_1}$ 
through the lead cross section.
Determination of transverse states $\chi_{m}(y_i)$ and of the corresponding 
longitudinal momentum $k_m$ as well as the normalization factors
$\theta_m$ will be discussed below.
For reflection amplitudes $r_{nm}$,  a  relation similar to (\ref{2.17})
holds.\cite{bara89} From $t_{nm}$ and $r_{nm}$ the conductance can 
be calculated by means of the Landauer formula [Eq.~(\ref{eq:landauer2})].
\subsection{Calculation of modules}
\label{sec:dyson}
This section is dedicated to the evaluation of 
the Green's functions for those modules, which we need to assemble
a circle and a stadium billiard: 
the half-infinite leads, the rectangle, the circle, and the half-circle. 
With 
the exception of the circle, for all these modules Eq.~(\ref{2.3}) is not 
applicable. This is due to the non-separability for confined 
geometries as discussed above. Moreover the spectrum in open 
structures like the semi-infinite lead 
is continuous rather than discrete. Unlike in the field-free 
case,\cite{econ79} the resulting integrals cannot be calculated 
analytically. However, both problems can be overcome by applying the matrix 
Dyson equation in a non-standard way.
\subsubsection{Rectangular module}
As illustrated above for the semi-infinite wave guide, the 
Dirichlet boundary condition for the confined structure of a 
rectangle with magnetic field is not separable, no 
matter which gauge is chosen. The separability can however
be restored by imposing periodic boundary 
conditions on two opposing sides of the rectangle. 
Topologically, this corresponds to folding the rectangle 
to the surface of a cylinder (Fig.~\ref{fig:dyson2}a). 
In this case we connect the 
first ($P$) and the last ($Q$) transverse grid slice of a
rectangular grid by a hopping potential 
$|V^x_{PQ}|=|V^x_{QP}|=\frac{-1}{2\Delta x^2}$.
The Green's function of this ``cylinder surface'' $(cs)$ 
will be denoted by $G^{\rm cs}$ in the following. 
The calculation of the rectangle Green's function $G^{\rm r}$ will be
obtained out of $G^{\rm cs}$ by a Dyson equation used here
in ``reversed'' mode, i.e.~for {\it disconnecting} tb grids.
This method for calculating the rectangular
module may seem like a detour, but it is numerically 
more efficient than a strip-by-strip recursion. 
For completeness we mention that an alternative way to calculate
$G^{\rm r}$ was proposed in Ref.~\onlinecite{skja94}.\\The Green's function 
for the cylinder surface $G^{\rm cs}$ 
can be constructed from separable eigenfunctions,
$|E_m\rangle=|E^{x}_{k}\rangle\otimes|E^y_{kn}\rangle$, according to 
Eq.~(\ref{2.3}).
Solving the tight-binding Schr\"odinger equation for the cylinder surface, we 
obtain for the longitudinal eigenstates
$\langle x_j|E_k^x\rangle=(N_x\Delta x)^{-1/2}\exp(i2\pi kj/N_x)$, which 
results in a tridiagonal, symmetric matrix-eigenproblem 
of size $N_y\times N_y$ for the transverse modes,\cite{numr86}
\begin{eqnarray}
E_{kn}\langle y_l|E^{y}_{kn}\rangle=\frac{-1}{\Delta x^2}\left[\cos\left(
\frac{2\pi k}{N_x}+\frac{B}{c}y_l\Delta x\right)-1\right]\nonumber\\\times
\langle y_l|E^{y}_{kn}\rangle
-\frac{1}{2\Delta y^2}\big(\langle y_{l-1}|E^{y}_{kn}\rangle-
2\langle y_l|E^{y}_{kn}\rangle+\langle y_{l+1}|E^{y}_{kn}\rangle\big).
\end{eqnarray}
By ``cutting the cylinder surface open'' 
along a line of constant $x$, we obtain from $G^{\rm cs}$ the 
desired  Green's function $G^{\rm r}$ for the rectangle 
(Fig.~\ref{fig:dyson2}a). 
We demonstrate this for  the rectangle Green's function $G_{PX}^{\rm r}$ from the first transverse slice $P$ to any other slice $X$.
To determine $G_{PX}^{\rm r}$ we solve the following system of 
Dyson equations,
\begin{eqnarray}\label{IIB9}
G_{PX}^{\rm r}&=&G_{PX}^{\rm cs}-G_{PQ}^{\rm r}\bar{V}_{QP}G_{PX}^{\rm cs}
-G_{PP}^{\rm r}\bar{V}_{PQ}G_{QX}^{\rm cs}\\
G_{PQ}^{\rm cs}&=&G_{PQ}^{\rm r}+G_{PQ}^{\rm r}\bar{V}_{QP}G_{PQ}^{\rm cs}
+G_{PP}^{\rm r}\bar{V}_{PQ}G_{QQ}^{\rm cs}\\
G_{PP}^{\rm cs}&=&G_{PP}^{\rm r}+G_{PQ}^{\rm r}\bar{V}_{QP}
G_{PP}^{\rm cs}+G_{PP}^{\rm r}\bar{V}_{PQ}G_{QP}^{\rm cs}\,,
\end{eqnarray}
where the first line is the ``reversed'' Dyson equation.
The three unknowns in the above equations are the  Green's functions 
connecting the slices $(P,X), (P,Q)$ and $(P,P),$
$G_{PX}^{\rm r},G_{PQ}^{\rm r},G_{PP}^{\rm r}$. 
By solving these three  equations, the unknowns can be uniquely determined.
\subsubsection{Circle and half-circle}
\begin{figure}[h] 
\includegraphics[draft=false,keepaspectratio=true,clip,width=0.65\linewidth]{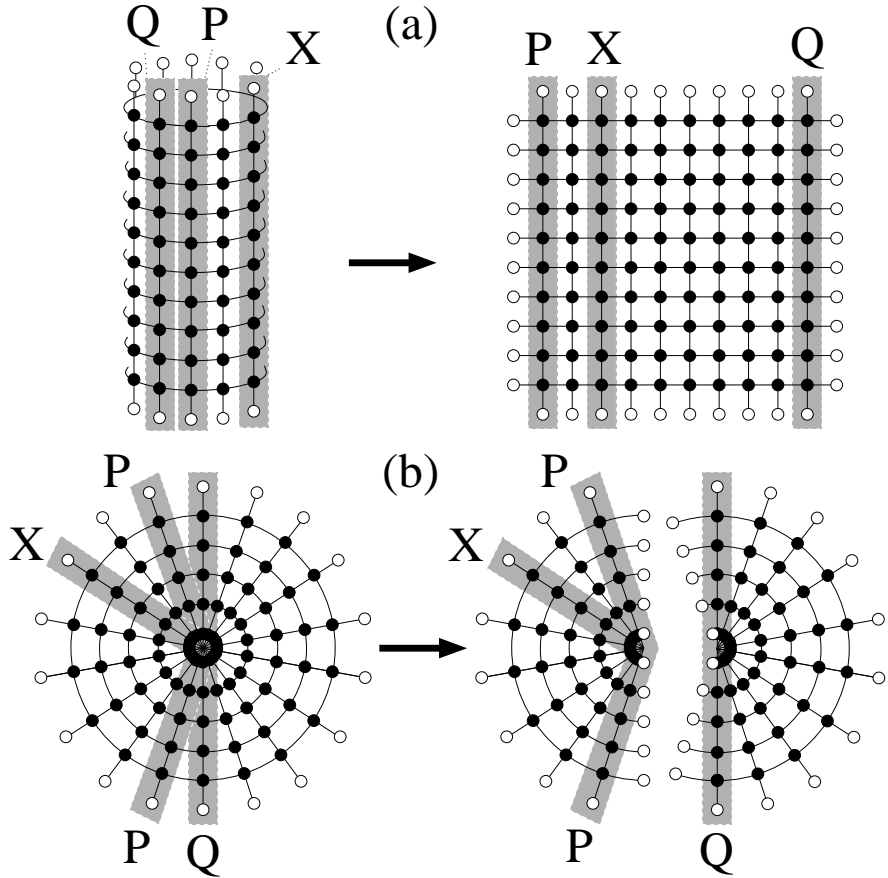}%
\caption{Applying a Dyson equation in ``reversed mode''  to construct
Green's functions for (a) a rectangle  out of 
a cylinder surface and (b) a semi-circle out of a full circle, 
respectively. In  (a) the periodic
boundary conditions are transformed 
into hard wall boundary conditions. 
The gray shaded areas $P$, $Q$ and
$X$ are those grid slices at which the Green's functions 
are evaluated (see text).}
\label{fig:dyson2}
\end{figure}

In symmetric gauge, ${\bf A}=B/2(-y,x,0)$, 
the Dirichlet boundary value problem for 
the circle with magnetic field is separable,
$|E_m\rangle=|E^{\varphi}_{k}\rangle\otimes|E^\varrho_{kn}\rangle$. 
On a discrete tb lattice this statement remains true, 
provided that a circular grid is employed. 
With the eigenstates for
the azimuthal degree of freedom, $\langle
\varphi_j|E_k^\varphi\rangle=
(N_\varphi\Delta\varphi)^{-1/2}\exp(i2\pi kj/N_\varphi)$ and 
radial eigenstates
$g_{kn}(\varrho_i)=\sqrt{\varrho_i}\times\langle\varrho_i|
E^{\varrho}_{kn}\rangle$, the finite difference equation for 
the $g_{kn}(\varrho_i)$ results in a tridiagonal symmetric 
eigenproblem
\begin{eqnarray}
   E_{kn}\,g_{kn}(\varrho_i)&=&-\frac{1}{\varrho_i^2\Delta\varphi^2}\left[
\cos\left(\frac{2k\pi}{N_\varphi}-\frac{\varrho_i^2B\Delta\varphi}{2c}
\right)-1\right]\,g_{kn}(\varrho_i)\nonumber\\&&-
\frac{1}{2\Delta\varrho^2}\left[
\frac{\varrho_{i-1/2}}{\sqrt{\varrho_{i-1}}\sqrt{\varrho_{i}}}
\,g_{kn}(\varrho_{i-1})-
2\,g_{kn}(\varrho_i)+\frac{\varrho_{i+1/2}}{\sqrt{\varrho_{i}}
\sqrt{\varrho_{i+1}}}
\,g_{kn}(\varrho_{i+1})\right]\,.
\end{eqnarray}
The Green's function for the circular module is then calculated
by a straight-forward application of  Eq.~(2.3).\\For 
the Green's function of the half-circle we
employ an analogous procedure as in the previous subsection: we
dissect the circle Green's function into half-circles by
means of a ``reversed'' Dyson equation.
We demonstrate this by way of the example depicted
in Fig.~\ref{fig:dyson2}b, 
where the ``full circle'' (fc) is split up into two ``half-circles'' 
(hc). The resulting two halves are almost identical, with the exception
of the two additional radial grid slices, by which the right half-circle is 
larger. For assembling the stadium billiard we have to make sure that the
tb grid of the half-circle module  can be linked directly
to a vertical grid  [see Fig.~\ref{fig:3}b]. 
For this reason, only the left one of the two half-circles 
in Fig.~\ref{fig:dyson2}b 
can be used for this purpose.\\Consider as example  the
Green's function $G_{PX}^{\rm hc}$ describing the propagation 
from the grid slice $P$  at the 
junction of the two half-circles to any radial grid slice $X$ situated
on the ``left half-circle'' (see Fig.~\ref{fig:dyson2}b).
$G_{PX}^{\rm hc}$ is determined by the following system of Dyson equations
\begin{eqnarray}
G_{PX}^{\rm hc}&=&G_{PX}^{\rm fc}-G_{PP}^{\rm hc}\bar{V}_{PQ}G_{QX}^{\rm fc}\\
G_{PP}^{\rm fc}&=&G_{PP}^{\rm hc}+G_{PP}^{\rm hc}\bar{V}_{PQ}G_{QP}^{\rm fc}
\end{eqnarray}
which yields a unique solution for $G^{\rm hc}_{PX}$.
\subsubsection{Semi-infinite lead} 
\begin{figure}[!tbh] 
\includegraphics[draft=false,keepaspectratio=true,clip,width=0.7\linewidth]{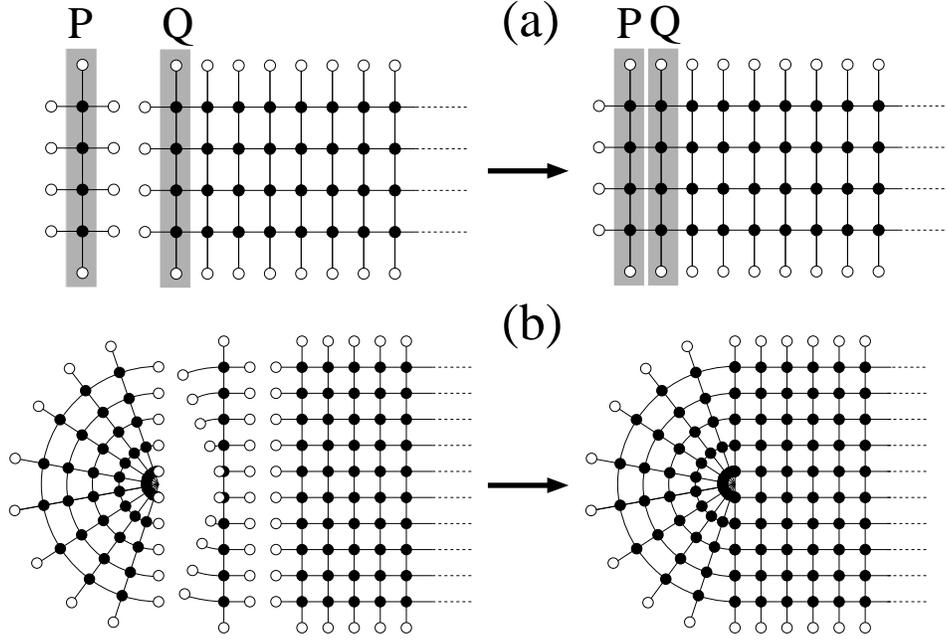}%
\caption{Applying a Dyson equation to construct
Green's functions for (a) a semi-infinite lead and 
(b) a stadium billiard out of ``modules''.
In (a) joining a transverse slice with a semi-infinite lead schematically 
leaves the Green's function of the lead invariant. In (b) 
an additional link module is added to facilitate the coupling
between the half-circle and the rectangle module.
Notation as in Figs.~\ref{fig:dyson1} 
and \ref{fig:dyson2}.}
\label{fig:3}
\end{figure}

Because of its continuous spectrum, the Green's function for the 
semi-infinite lead poses an additional challenge beyond that of 
the non-separability of the wavefunction discussed above. We  
therefore apply one further ``trick'' to bypass this problem. 
Our approach is based on the observation that adding a slice to 
a semi-infinite quantum wire leaves this wire (up to 
irrelevant phases) invariant (see Fig.~\ref{fig:3}a). We assume a 
semi-infinite lead with
$x\in[\Delta x,\infty)$ and hard-wall boundary 
conditions at $x=\Delta x$ and $y=\pm d/2$. To this object we add a 
slice consisting of just one transverse chain of tb grid 
points which we place at $x=0$.
The system of  Green's functions for the propagation  
from the transverse chain at $x=0$ ($P$) back to itself $(P)$ or to 
the first transverse slice of the semi-infinite 
lead ($Q$) at $\Delta x$ reads  
\begin{eqnarray}\label{2.25}
G_{PP}=G^0_{PP}+G^0_{PP}\bar{V}_{PQ}G_{QP},\\
G_{QP}=G^0_{QQ}\bar{V}_{QP}G_{PP}.\label{2.26}
\end{eqnarray}
Each multiplication involves a matrix product with a dimension equal 
to the number of transverse grid points. The key point is now 
that the system of Eqs.~(\ref{2.25},\ref{2.26}) can be closed 
through the invariance condition (Fig.~\ref{fig:3}a) for the semi-infinite 
lead, i.e.~$G_{PP} = G^0_{QQ}$. In Landau gauge 
${\bf A}=(-By,0,0)$ the latter relation does not involve additional 
gauge phases since these are already 
contained in the hopping matrix element. We additionally note that
an equivalent point of departure for the derivation of $G_{PP}$ 
is the Bloch condition for states in the 
lead.\cite{ando91,zozo96}\\Setting $Z=G_{PP} 
\bar{V}_{QP}$ and using the hermiticity 
condition $\bar{V}_{QP}=\bar{V}^*_{PQ}\equiv\bar{V}^*$, 
Eqs.~(\ref{2.25},\ref{2.26}) can be converted to a quadratic matrix equation 
\begin{equation}\label{2.27}
Z Z-\bar{V}^{-1}(G^0_{PP})^{-1}Z+\bar{V}^{-1}\bar{V}^*=0\,.
\end{equation}
Solvents $Z$ of a quadratic matrix
equation $Q(Z)=0\,$ can be constructed from the eigenpairs $(\beta_i,
{\chi}_i)$ 
of the corresponding quadratic eigenvalue equation 
$Q(\beta_i){\chi}_i=0,\, i\in[1,\ldots,2N]$ in the diagonal 
form,\cite{high00} 
\begin{equation}\label{IIB2}
Z=MB M^{-1}\quad {\rm with} \quad M=[{\chi}_1,\ldots,
{\chi}_N],\,
B={\rm diag}(\beta_i).
\end{equation} 
The quadratic eigenvalue equation is equivalent to
a generalized eigenvalue problem $A\tilde{\chi}=\beta C\tilde{\chi}$
of twice the original dimension.\cite{numr86} Its $2N$ dimensional eigenvectors
$\tilde{\chi}=({\chi},\beta{\chi})$ are solutions of the 
symmetric eigenproblem
\begin{equation}\label{2.29}
\left(\begin{array}{cc}-\bar{V}^*&0\\0&\bar{V}\end{array}\right)
\left(\begin{array}{c}{\chi}\\\beta {\chi}\end{array}\right)=\beta
\left(\begin{array}{cc}-(G^0_{PP})^{-1}&\bar{V}\\\bar{V}&0\end{array}\right)
\left(\begin{array}{c}{\chi}\\\beta{\chi}\end{array}\right)\,,
\end{equation} 
where $\left( G^0_{PP} \right)^{-1}=E_F-\hat{H}^{\rm tb}_{\rm 1D}$ and 
$\hat{H}^{\rm tb}_{\rm 1D}$ is the Hamiltionian of the one-dimensional 
transverse tb strip at $x=0$. The Fermi energy
$E_F$ and the magnetic field $B$ enter (\ref{2.29}) as 
independent parameters at which the eigenstates $\tilde{\chi}_m$ and 
eigenvalues $\beta_m$ are evaluated. The longitudinal momenta
of the lead states $\xi_m(x,y)=\chi_m(y)e^{ik_mx}/\sqrt{\theta_m}$
are related to the eigenvalues by the relation $\beta=\exp(ik\Delta x)$. 
The orthonormalisation and the completeness relations 
of the $2N$ eigenvectors $\tilde{\chi}_m$
can be formulated in terms of matrix relations, for the 
generalized eigenproblem, 
\begin{equation}\label{2.30}
\frac{1}{\sqrt{\theta_m\theta_n}}
\tilde{\chi}_m^T C\tilde{\chi}_n
=2i\frac{k_m}{|k_m|}\delta_{mn}
\quad{\rm and}\quad\sum_m^{2N} \frac{\tilde{\chi}_m 
\tilde{\chi}_m^T}{\theta_m}=2i\frac{k_m}{|k_m|}C^{-1}.
\end{equation}
With this specific choice of normalization the norm
factors $\theta_m$ are determined such that every
propagating state carries unit flux.
We note parenthetically that the quadratic eigenvalue equation 
could also be applied to the semi-infinite lead at zero $B$ field. 
However, in that case, the Green's function for quantum wires can 
be calculated analytically\cite{rott00,econ79} by
complex contour integration.
\subsubsection{Scattering wave functions and efficiency of the MRGM} 
The MRGM is particularly well-suited to  determine transport 
coefficients as the Green's function is then required only at the 
junctions between the modules and does not have to be evaluated
throughout the interior of the entire quantum dot. Since for the 
calculation of the scattering wavefunction the Green's 
function throughout the entire scattering region is needed, 
this particular advantage cannot be made use of here. 
However, also in this case, 
the MRGM is still more efficient than the standard RGM, as will be
explained below.\\The wavefunction  $\psi({\bf x})$ can be obtained
at any point ${\bf x}$ by projecting the retarded
Green's function (by means of the operator 
$\mathord{\buildrel{\lower3pt\hbox{$\scriptscriptstyle
\leftrightarrow$}}\over {\bf D}}$) on the incoming wave 
(in mode m),\cite{bara89,zozo96}
\begin{equation}\label{2.31}
\psi_m({\bf x})=-\frac{1}{2\sqrt{\theta_m}}\int^{d/2}_{-d/2} dy'_1
G^+({\bf x},{\bf x'},E_F,B)\,
(\mathord{\buildrel{\lower3pt\hbox{$\scriptscriptstyle
\leftrightarrow$}}\over {\bf D}}{\phantom{.}\!'}\cdot\hat{\mathbf x}'_1)\,
\chi_{m}(y'_1)e^{ik_{m}x'_1}\,.
\end{equation}
$G^+$ contains the solution of the Dyson equations for all 
linked modules. That the evaluation of Eq.~(\ref{2.31}) can be done
very efficiently results from 
two properties: First the number of recursions (i.e., of matrix 
inversions) needed to obtain $G^+$ 
is given by the fixed number of modules required 
to build up the scattering structure. This number is independent 
of the De Broglie wavelength. The latter enters only in terms of 
the size of the matrices involved in the recursion, 
since with increasing $E_F$ 
(decreasing $\lambda_{D}$) more grid points are required to 
represent the continuum limit. Compared to the standard RGM the 
numerical effort is reduced since in that approach the number of 
recursions scales with the grid density, i.e.~$\propto k_F$. A second 
advantage of the MRGM concerns the incorporation of the boundary 
conditions. In the modular method the boundaries follow by construction 
the nodal lines of the symmetry-adapted coordinate system for the 
module. Due to this reason the convergence towards the continuum 
limit is enhanced as compared to the slice-by-slice recursion. The 
calculation of the transport coefficients as a function of the 
Fermi wavenumber $k_F$ (or Fermi energy $E_F$) is simplified by 
the fact that the solution of the eigenvalue problem 
($|E_m\rangle, E_m$) entering the Green's function for each 
module [Eq.~(\ref{2.3})] is independent of $E_F$. For the 
evaluation of the Green's function at different values of 
$E_F$ the eigenproblem 
$\hat{H}^{\rm tb}|E_m\rangle=E_m|E_m\rangle$ therefore has 
to be solved
only once. Unfortunately, this feature does not extend to 
the variation of the magnetic field since both  
$|E_m\rangle$ and $E_m$ are  dependent on $B$. Because of this
property a new solution of the tb eigenproblem 
is required for each value of the field. The 
most severe restriction of 
the MRGM is, however, that its applicability is limited to those 
scattering structures which can be assembled from or cut out of separable
modules. Also random potentials and  soft walls can only be included 
as long as they preserve the separability of each module. 
We mention at this point, that  
a ``hybrid RGM'' for dealing with arbitrary boundary geometries 
was presented in the literature.\cite{zozo96}
\section{Numerical results for high $B$ fields and large $k_F$}
In this section we present first magnetoconductance results
which were calculated within the MRGM at 
high magnetic fields $B$ and large Fermi wavenumbers $k_F$. 
As prototype shapes of the cavity we use the circle and the 
Bunimovich stadium and consider different
geometries for the attached quantum wires. 
\subsection{Accuracy checks}
\begin{figure}[!tbh] 
\includegraphics[draft=false,keepaspectratio=true,clip,width=0.5\linewidth]{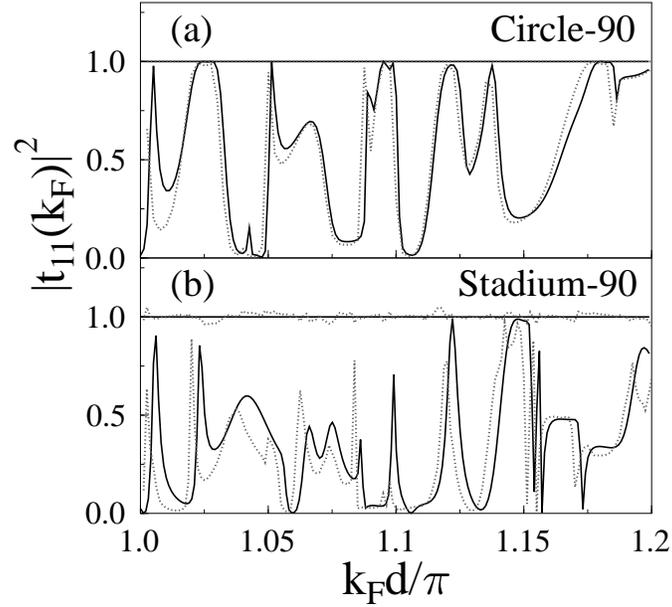}%
\caption{Comparison between the present MRGM (solid line) and the wavefunction
matching technique\cite{yang95} (dotted line) for the first-mode transmission
probability $|t_{11}(k_F)|^2$ at $B/c=1$ in a small window of $k_F$: 
(a) circle with
perpendicular leads, (b) stadium with perpendicular leads
($d=0.35,A^{\rm dot}=4+\pi$).
In both cases also $|t_{11}(k_F)|^2+|r_{11}(k_F)|^2$ is shown. 
Contrary to the MRGM (solid line), the wave function
matching technique (dotted line) deviates  from the unitarity
limit in (b).}
\label{fig:yang}
\end{figure}

Several checks for the accuracy of the numerical results have been 
performed. Exact relationships for transport coefficients such as 
conservation of unitarity and the Onsager relations are fulfilled 
with an accuracy of better than $10^{-10}$. The grid density is 
chosen such that the magnetic flux per unit cell is $B \Delta_r/c < 0.01$
(as in Ref.~\onlinecite{bara91}). 
Moreover, the typical number of grid points per Fermi 
half-wavelength is greater than $30$. Only for very high energy calculations
(Fig.~\ref{fig:highenergy}) the relative grid density is lower. For 
low magnetic fields, we can compare our 
results for $|t_{nm}(k_F)|^2$ with
previous methods. As an example we show in Fig.~\ref{fig:yang} a 
comparison for $|t_{11}(k_F)|^2$ 
with the calculation by Yang {\it et al.},\cite{yang95} 
which is based on a wave function 
expansion in spherical waves. The agreement for the circle is very good
although diamagnetic terms are neglected in the approach of 
Ref.~\onlinecite{yang95}. For
the stadium, the differences between the two methods are somewhat larger.
This is due to the fact that the expansion of the stadium wave functions
in spherical waves leads to a unitarity deficiency 
(see Fig.~\ref{fig:yang}). We can also reproduce previous 
results of Ref.~\onlinecite{yang95} concerning
weak-localization line shapes and 
statistical magnetoconductance properties in chaotic and regular cavities.
These will not be treated again. Our focus will be  
on the high magnetic field and high energy regime, 
where other methods failed.
\subsection{Wavefunctions}
The starting point for the analysis of the scattering states 
$\psi {({\bf x})}$ for ballistic transport through quantum dots is
Eq.~(\ref{2.31}). Figures \ref{fig:highenergy} and \ref{fig:edgepics} 
display the resulting electron 
density $\propto |\psi({\bf r})|^2$ in the scattering region. In 
Fig.~\ref{fig:highenergy}  we 
consider the wavefunctions at very high $k_F$ for both the 
circle and the stadium billiard, which are prototypical structures for 
regular and chaotic dynamics respectively. Large $k_F$ corresponds to 
the regime where the convergence towards classical scattering 
trajectories is expected to emerge. 
Figures \ref{fig:highenergy}a and 
\ref{fig:highenergy}b illustrate the different 
dynamics for an injection at high ($m=20$) and at low 
mode number ($m=1$), respectively. Since high mode numbers correspond 
classically to a large injection angle, the wavefunction condenses 
around a pentagon-shaped whispering gallery trajectory. For 
low-mode injection, a small circle representing the centrifugal 
barrier (or caustic) is seen, as well as rays representing the asterisk 
orbits.\cite{ishi95} Figures \ref{fig:highenergy}c and 
\ref{fig:highenergy}d display scattering states 
for the stadium. At low magnetic fields, the dynamics is 
chaotic and a typical wavefunction features a quasi-random 
pattern with a modest density enhancement near classically 
unstable periodic orbits (not shown). For special configurations of
$k_F$ and $B$ ``scars'' emerge in the scattering wavefunctions 
(Fig.~\ref{fig:highenergy}c). By contrast, for high 
magnetic fields the classical motion becomes regular. In 
the present example the wavefunction condenses around a ``bundle''
of cyclotron orbits executing three bounces at the cavity 
wall before exiting by the entrance lead 
(Fig.~\ref{fig:highenergy}d). There has been an extensive discussion 
in the literature as to the existence of scars in open quantum 
billiards.\cite{akis97} Our present results clearly underscore 
that  scars, defined here as the condensation of the wavefunction 
near classical (not necessarily unstable) trajectories, 
clearly exist for large $k_F$. Figure \ref{fig:edgepics} 
illustrates the formation 
of edge states at high fields. With increasing $B$ fewer edge 
states can be excited in the cavity. 
In Fig.~\ref{fig:edgepics}c ($B=68.5$) three transverse edge states 
are present while in Fig.~\ref{fig:edgepics}d ($B=125$) 
only a single edge state remains. 
For two edge states carrying flux across the quantum dot,
interferences give rise to a stationary nodal pattern 
with a fixed number of antinodes along the boundary (see 
Fig.~\ref{fig:edgepics}a,b). We note that
we are not aware of any 
other method that has so far been capable of investigating 
scattering states of open structures in this high-magnetic field 
regime.
\subsection{Transport coefficients}
\begin{figure}[!tbh] 
\includegraphics[draft=false,keepaspectratio=true,clip,width=0.8\linewidth]{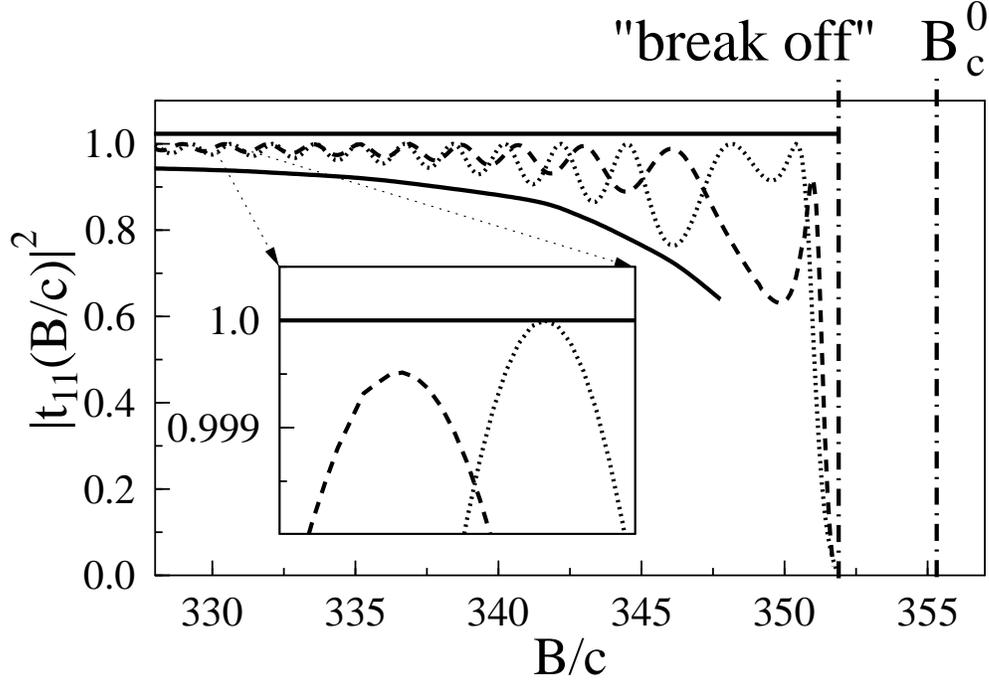}%
\caption{Transmission probabilities $|t_{11}(B/c)|^2$
in the high-field limit, near the point where transport terminates.
The dotted line stands for the circle billiards in both
lead geometries (their transmission probabilities are identical) 
and the dashed line for the stadium billiard
with $90^\circ$ lead geometry. The solid curves represent the 
upper and lower bounds of the oscillations
(offset for better visibility).
The two dash-dotted vertical lines mark the point where 
transport breaks off
and the analytically determined threshold value $B_c^0\approx 355.3$
(see text for details).
The inset shows that the  transmission probabilities for the circle
reach the maximum value 1 which is only approximately true for the stadium
with $90^\circ$-lead geometry.}
\label{fig:spectra2}
\end{figure}

The interference  between different edge states gives rise to 
characteristic fluctuations in the transport coefficients.
Figure \ref{fig:spectra} shows the high magnetic field
regime of the transmission probability in the first mode 
$m=n=1$ for both circle and stadium. Different orientations 
of the exiting quantum wire were chosen (oriented $90^\circ$ and 
$180^\circ$ relative to 
the incoming lead). A few universal trends are easily 
discernible: above a certain critical value of the magnetic 
field (denoted by $B^1_c$), the strongly fluctuating 
transmission probability gives way to very regular oscillations  
in all four cases [see insets of Fig.~\ref{fig:spectra} for magnification]. 
The threshold value $B^1_c$ and the magnetic field, at which 
transport is terminated (separately displayed in Fig.~\ref{fig:spectra2}) are 
identical for all systems investigated. Below $B^1_c$ the 
transport signal displays ``beats'', i.e.~the Fourier transform 
of the signal is characterized by several frequencies. The
``universality'' (i.e.~geometry independence) of these features is 
related to the fact that in the high magnetic field regime 
transport is controlled by edge states (as depicted in 
Fig.~\ref{fig:edgepics}). 
These states play a very prominent role in the Quantum Hall effect 
and have been studied 
extensively.\cite{pran87,ferr97,davi98,datt95} At 
magnetic fields, where the magnetic length is smaller than
the system dimensions, $l_B\ll D$, they are the only  states 
coupling to the quantum wire since
bulk Landau states cannot be excited by the leads.
The edge states shown in Figs.~\ref{fig:edgepics}a to 
\ref{fig:edgepics}d  correspond
to the points in the transmission spectrum also labeled by (a) to (d) 
in Fig.~\ref{fig:spectra}.
By comparison with the scattering wavefunctions (as in
Fig.~\ref{fig:edgepics}), we observe that in the magnetic field 
region $B_c^n<B<B_c^{n-1}$
edge states have up to $n-1$ transverse nodes in the direction 
perpendicular to the boundary. Furthermore, the number of longitudinal 
antinodes from entrance to exit lead (see the corresponding numbers
in Fig.~\ref{fig:edgepics}a,b) can be directly mapped onto 
successive maxima in Fig.~\ref{fig:spectra} (see numbers there). 
The range of $B$ depicted in Fig.~\ref{fig:spectra} 
corresponds to $B\gtrsim B^2_c$ at
$k_F=1.5\pi/d$. The transmission spectrum 
becomes increasingly complex as $B$ is 
reduced or equivalently $k_F$ is increased (not shown).\\To 
determine the positions of the values $B^n_c$ 
we consider the energy shift of  Landau levels
near the boundary. Bulk Landau levels are degenerate since
their quantized energy $E_n=(n+1/2)B/c$ is independent of their
positition in space. This degeneracy is lifted if a Landau state
is placed in the vicinity of the cavity wall: with decreasing distance 
to this boundary the energy of the state increases. Therefore
the energies of edge states 
associated with the quantum number $n$ lie above the asymptotic
bulk value $E_n$. When the incoming electron is diffracted at the 
mouth of the entrance lead, only those edge states whose energy is 
below the Fermi energy can carry the flux. Due to the sharp edges at 
the junction between lead and quantum dot, all energetically 
accessible edge channels are populated. The blow-up of the scattering 
wavefunction near the lead mouth (Fig.~\ref{fig:labelnew}) 
highlights the diffractive 
edge scattering. This is a contrast to smooth edges where states in 
the lead could cross the lead junction adiabatically, i.e.~without 
changing their state of quantization.\cite{glaz90} With sharp lead
junctions however all edge states 
with quantum number $n'\le n$ will participate in transport up to a 
magnetic field where $E_n$ touches the Fermi energy, $E_n=E_F$, i.e.~at 
the critical magnetic fields $B_c^n/c=E_F/(n+1/2)$. These 
threshold values are indicated by the dot-dashed vertical bars in 
Fig.~\ref{fig:spectra},\ref{fig:spectra2} for
 $B_c^2\approx 71.1$, $B_c^1\approx 118.4$ and $B_c^0\approx 355.3$
for a lead width
$d=0.25$ and $k_F=1.5\pi/d$. In our numerical data, both 
the position of these threshold 
values as well as their independence of the geometry are in excellent 
agreement with this prediction. The only exception is the 
critical magnetic field $B_c^0$. Its value ($355.3$) lies slightly 
above the point where the transmission spectrum ceases (at $B\approx 351.8$) 
(see Fig.~\ref{fig:spectra2}). 
The reason for this deviation is the fact
that the termination point of the spectrum is not determined by the 
magnetic field of the lowest bulk Landau level in the cavity 
(i.e.~$B_c^0$), but by the highest field at which the leads
still carry flux. In the leads, however, the magnetic length does not 
satisfy the condition $l_B\ll d$ (at $k_F=1.5\pi/d: l_B\approx d/4.7$).
Contrary to the cavity, the lead wavefunctions still ``feel'' the
constriction by the opposing wall. For this reason the 
threshold magnetic field values of the transverse lead states 
lie slightly below those of the bulk Landau levels, leading to a termination
already below $B_c^0$.
\subsection{Multi-channel interferences}
\begin{figure}[!tbh] 
\includegraphics[draft=false,keepaspectratio=true,clip,width=0.8\linewidth]{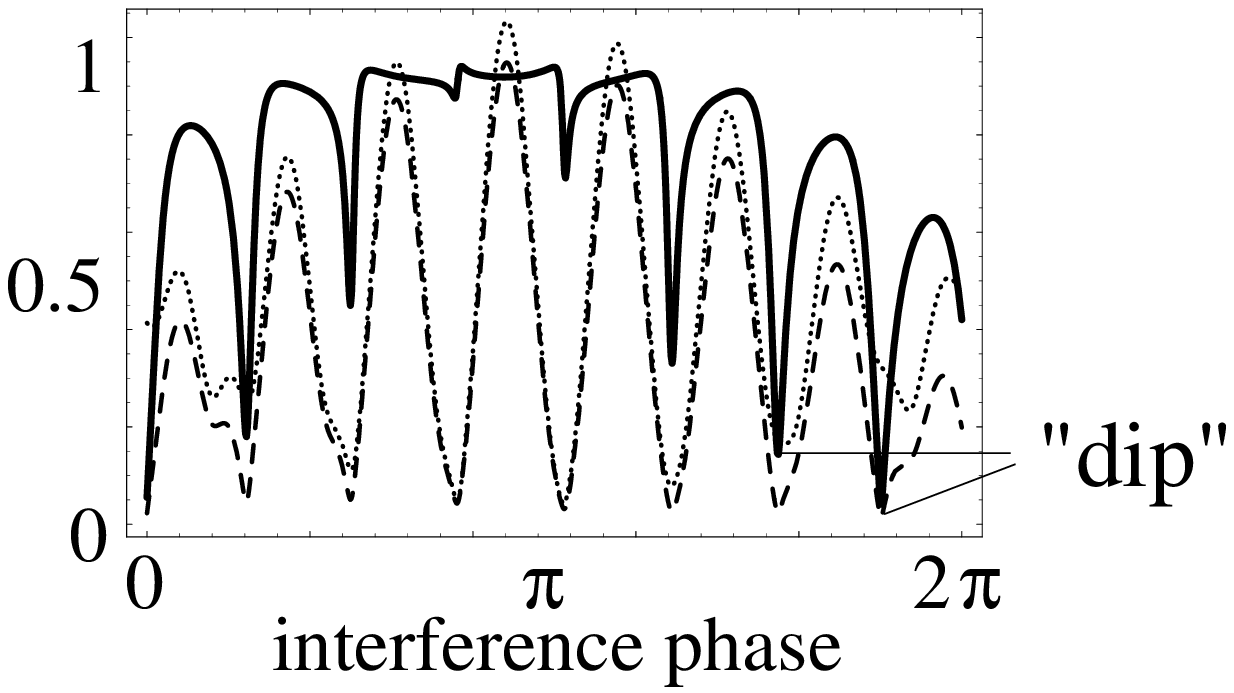}%
\caption{One half-period of the beating
$\pi n <B(A_1-A_2)/(2c)<\pi(n+1)$ 
in the transmission
probability $|t_{11}(B/c)|^2$ (solid line), 
as calculated with our interference model [see Eq.~(\ref{eq:fano})]. 
The nominator (dashed line, $N$) and denominator (dotted line, $D$)
of $|t_{11}(B/c)|^2$
show very similar oscillations (with a small offset). ($N/D$) 
features sharp ``dips'', 
at the points where $N$ and $D$ have their common minima.
These dips are \textit{window resonances} (also called 
Breit-Wigner antiresonances) and represent a symmetric 
limit of the Fano resonance lineshape. 
See text for details.}
\label{fig:spectra3}
\end{figure}

The regular oscillations above $B^1_c$ as well as the complex 
fluctuating pattern below $B^1_c$ can be explained by a multi-channel 
scattering description. This model can be viewed as a 
generalization\cite{kim01,wirt02} of a 
single-channel picture.\cite{siva89,wees89} For this description to be
applicable, the cavity of the dot has to have smooth boundaries and
disorder must be absent. Under these circumstances the flux 
transported by edge states is conserved in the interior and changed 
only by diffractive scattering at lead junctions: At the junction, 
a fraction of the flux will exit through the lead while the remaining 
portion of the flux will continue to propagate along the 
boundary. The stationary scattering state can be viewed as the 
coherent superposition of repeated loops around the billiard. In 
order to translate this picture into an analytic expression we 
define amplitudes for
transmission and reflection at the two  lead junctions. We denote
the amplitudes for transmission from transverse 
mode $m$ in the 
entrance lead to the edge state in the dot with quantum 
number $i$  by $\tilde{t}_{mi}$.
The amplitudes $\tilde{t}'_{in}$ stand for transmission from edge state 
$i$ in the dot
to the transverse mode $n$ in the exit lead. The amplitudes 
$\tilde{r}_{ij}(\tilde{r}'_{ij})$ describe edge state reflection
at the entrance (exit) lead from mode $i$ to mode $j$. (The tilde signs 
serve to distinguish these amplitudes from the transport coefficients
for the whole geometry $t_{nm}$ and $r_{nm}$.) 
We further define  the following matrices
\begin{eqnarray}
{[\tilde{T}]}_{ij}&=&\tilde{t}_{ij}e^{ik_jL_j-iBA_j/c},\quad
{[\tilde{T}']}_{ij}=\tilde{t}'_{ij}\\
{[\tilde{R}']}_{ij}&=&\tilde{r}'_{ij}e^{ik_jL'_j-iBA'_j/c},\quad
{[\tilde{R}]}_{ij}=\tilde{r}_{ij}e^{ik_jL_j-iBA_j/c}\,,
\end{eqnarray}
where $L_j,A_j$ ($L'_j,A'_j$) denote the lengths $L$ and areas $A$ 
the edge state $j$
covers from entrance to exit (from exit to entrance) of the dot.
The areas $A$ 
can be determined in gauge-invariant form, although
the corresponding classical orbits are not necessarily
periodic.\cite{wirt99}
The transmission through the whole cavity $t_{ji}=[T]_{ij}$ is then
written as a geometric series of matrices, 
\begin{eqnarray}\label{3.3}
T&=&\tilde{T}(1+\tilde{R}'\tilde{R}(1+\tilde{R}'\tilde{R}
(1+\ldots)))\,\tilde{T}'\nonumber\\
&=&\tilde{T}\left(\sum_{i=0}^\infty(\tilde{R}'\tilde{R})^i\right)\tilde{T}'
\nonumber\\
&=&\tilde{T}(1-\tilde{R}'\tilde{R})^{-1}\tilde{T}'\,.
\end{eqnarray}
Equation (\ref{3.3}) serves as a convenient starting point for the 
analysis of the transmission fluctuations. Consider first the 
regime $B >B^1_c$, where only the lowest transverse edge state is 
excited. In this case Eq.~(\ref{3.3}) reduces to its scalar 
version\cite{siva89,wees89}
\begin{equation}\label{3.4}
T^{\rm tot}=|t_{11}|^2=\frac{|\tilde{t}_{11}|^2|\tilde{t}'_{11}|^2}
{1-2{\rm Re}\left[\tilde{r}'_{11}\tilde{r}_{11}
e^{i\gamma}\right]
+|\tilde{r}'_{11}|^2|\tilde{r}_{11}|^2}\,,
\end{equation}
with $\gamma=k_1(L_1+L'_1)-B(A_1+A'_1)/c$.
As expected, the fluctuations of $|t_{11}(B/c)|^2$ 
are determined by an Aharonov-Bohm type 
phase $\gamma$. At fixed $k_F$,
the oscillation period is $\Delta B=2\pi c/A_1^{\rm tot}$. By 
$A_1^{\rm tot}=A_1+A'_1$ we denote the area which the edge state
acquires in one revolution around the dot.
Taking into account that the dynamically accessible area of the edge 
state is somewhat smaller than the
geometric area, $A_1^{\rm tot}<A^{\rm dot}=4+\pi$ 
(see Fig.~\ref{fig:edgepics}), the prediction for the 
oscillation period is in excellent agreement with our numerical 
findings. Equation (\ref{3.4}) also explains 
why the oscillation period is increasing with increasing $B$ 
(see Fig.~\ref{fig:spectra2}). This explanation makes 
use of the somewhat 
counterintuitive fact, that for increasing magnetic field  
skipping orbits with fixed quantum number $n$
have an increasing mean distance from the boundary.\cite{wees89}
Consequently, a larger $B$ field implies a smaller enclosed area 
$A^{\rm tot}_1$ and therefore a higher oscillation period $\Delta B$. 
Furthermore Eq.~(\ref{3.4}) accounts for the fact that for most 
structures the successive maxima of $T^{\rm tot}$ reach unity.\cite{siva89} 
(The small deviation from this rule of the stadium 
with $90^\circ$ lead orientaion 
will be explained below). In addition to unitarity, 
$[|\widetilde{t}'_{11}|^2+|\widetilde{r}'_{11}|^2=1]$ we have 
for identical lead junctions $\widetilde{r}_{11}=\widetilde{r}'_{11}$. 
(We call two junctions \textit{identical} if the local environment
of their lead mouths is the same 
and their respective distance is larger than a few wavelenths.) 
Above $B_c^1$ scattering of an edge state
at a junction is essentially a one-dimensional process, for which the 
probability for transmission from left to right has the same magnitude 
as vice versa. Identical lead junctions therefore also 
imply $\tilde{t}'_{11}=\tilde{t}_{11}$, provided that the
two corners of the lead junction have the same shape.
If and only if all of the three above conditions are fulfilled, 
Eq.~(\ref{3.4}) yields $T^{\rm tot}=1$ at the resonance 
condition $\gamma=2\pi n,n\in{\mathbb Z}$.
Since for the two circle geometries and for the $180^\circ$-stadium 
the two lead junctions are identical, we indeed find in these
cases that $|t_{11}(B)|^2$ periodically reaches unity.
On the other hand, when the leads are attached to the stadium at 
an angle of $90^\circ$, one lead is attached to the straight section 
while the other is attached to the semicircle. The local environment 
of the two lead mouths is in this case different (i.e.~the lead junctions
are not \textit{identical}), for which reason our numerical results 
do not quite reach  $|t_{11}|^2=1$, when the resonance
condition is fulfilled for this geometry (see inset 
of Fig.~\ref{fig:spectra}d and Fig.~\ref{fig:spectra2}). 
Another relation exists
between the resonance condition and the behaviour 
of edge states. In the closed cavity 
an edge channel always encloses
an integer number of flux quanta [$BA/(\phi_0c)=m\in{\mathbb N}$]. 
Therefore, the resonance condition
is met whenever the energy of an edge state in the closed
cavity crosses the Fermi 
edge.\cite{wees89}\\One interesting feature of the transmission 
fluctuations in the single-channel regime of the 
circular dot ($B >B^1_c$) is their 
invariance  with respect to the lead orientation. 
The numerical results for the transmission probabilities of the
circle with $180^\circ$ and $90^\circ$ lead orientation
differ only at the tenth (!) decimal digit.
This fact, as well as the observation, that in the case of the stadium 
billard the two lead orientations give different results, 
can again be explained by Eq.~(\ref{3.4}). The important point
to note is that the  interference phase 
($\gamma=k_1L_1^{\rm tot}-BA_1^{\rm tot}/c$) does not 
change  when changing the positions of the leads around the circle.
Due to the rotational symmetry  also the 
coefficients $\tilde{t}_{11},\tilde{t}'_{11},\tilde{r}_{11}$ and
$\tilde{r}'_{11}$ are the same for different angles between the leads.
The same is thus true for the total transmission $T^{\rm tot}$ through
the circular dot. The only exception to this rule occurs when
the two leads are in close proximity to within a few wavelengths. 
In this case the transmission probability changes as compared to the
results for the $180^\circ$ and $90^\circ$ lead orientation 
(not shown).\\The fluctuations in the regime $B<B^1_c$ displayed in 
Fig.~\ref{fig:spectra} 
can be analyzed with the help of a multi-channel scattering 
description. In the interval $B^2_c \leq B \leq B^1_c$ two 
channels corresponding to two edge states are open in the cavity and
one channel in each of the leads.
From the entrance to the exit lead mouth the
two edge channels acquire the phases $e^{ik_1L_1-iBA_1/c}$ and 
$e^{ik_2L_2-iBA_2/c}$ 
respectively. Their interference at the exit lead will 
therefore give an oscillatory contribution to the total 
transmission $T^{\rm tot}(B)=|t_{11}|^2$ of the form 
$\cos^2[B(A_1-A_2)/(2c)]$. 
For closer analysis we need to evaluate
Eq.~(\ref{3.3}) which involves the inversion 
of $2\times2$ matrices. 
In the case of parallel lead orientation the corresponding
expressions are simplified due to the fact that the phases acquired from 
entrance to exit lead and vice versa are the same ($A_n=A'_n$ and $L_n=L'_n$), 
\begin{eqnarray}\label{3.5}
t_{11}&=&\left[e^{i\varphi_1}\tilde{t}_{11}\tilde{t}'_{11}
+e^{i\varphi_2}\tilde{t}_{12}\tilde{t}'_{21}
+e^{i(2\varphi_1+\varphi_2)}(\tilde{r}_{11}\tilde{t}_{12}-
\tilde{r}_{12}\tilde{t}_{11})
(\tilde{r}'_{21}\tilde{t}'_{11}-\tilde{r}'_{11}\tilde{t}'_{21})
\right.\nonumber\\
&&\left.+e^{i(\varphi_1+2\varphi_2)}(\tilde{r}_{21}
\tilde{t}_{12}-\tilde{r}_{22}\tilde{t}_{11})
(\tilde{r}'_{22}\tilde{t}'_{11}-\tilde{r}'_{12}\tilde{t}'_{21})\right]/
\left[1-e^{i2\varphi_1}\tilde{r}_{11}\tilde{r}'_{11}
-e^{i2\varphi_2}\tilde{r}_{22}\tilde{r}'_{22}\right.\nonumber\\
&&\left.-e^{i(\varphi_1+\varphi_2)}(\tilde{r}_{21}\tilde{r}'_{12}
+\tilde{r}_{12}\tilde{r}'_{21})
-e^{i2(\varphi_1+\varphi_2)}(\tilde{r}_{11}\tilde{r}_{22}-
\tilde{r}_{12}\tilde{r}_{21})
(\tilde{r}'_{12}\tilde{r}'_{21}-\tilde{r}'_{11}
\tilde{r}'_{22})\right]\,,
\end{eqnarray}
with the abbreviated notation $\varphi_n=k_nL_n-BA_n/c$.
In Fig.~\ref{fig:spectra3} we show
one half-period of the beats in $T^{\rm tot}(B)=|t_{11}(B)|^2$ for 
[$\pi n <B(A_1-A_2)/(2c)<\pi(n+1)$], 
as calculated with Eq.~(\ref{3.5}). 
The absolute square of the numerator (dashed line, $N$) 
and denominator 
(dotted line $D$) of Eq.~(\ref{3.5}) display  
very similar oscillations, both in frequency and amplitude. However,
since $T^{\rm tot}=N/D$, a series of dips are superimposed on the
term $\cos^2[B(A_1-A_2)/(2c)]$ at the points 
where $N$ and $D$ have their common minima. To classify
these dips (i.e.~antiresonances) we make use of the
fact,\cite{kim01} that the unitarity of Eq.~(\ref{3.5})
can be satisfied by mapping the transport coefficients at the
lead junctions (which are assumed to be identical) 
onto six independent parameters 
$(r,\epsilon,\phi,\vartheta,\phi_1,\phi_2)$,
\begin{eqnarray}\label{kim1}
&\tilde{t}_{11}=\tilde{t}'_{11}=
\sqrt{(1-r^2)\epsilon}\,e^{i[(\phi_1+\psi)/2+\vartheta]},&\nonumber\\
&\tilde{t}_{12}=\tilde{t}'_{21}=
\sqrt{(1-r^2)(1-\epsilon)}\,e^{i[(\phi_2+\psi)/2+\vartheta]},&\nonumber\\
&\tilde{r}_{11}=\tilde{r}'_{11}=-[(1-\epsilon)+\epsilon r]
\,e^{i(\phi_1+\vartheta)},&\\
&\tilde{r}_{12}=\tilde{r}'_{12}=\tilde{r}_{21}=\tilde{r}'_{21}=
(1-r)\sqrt{\epsilon(1-\epsilon)}\,e^{i[(\phi_1+\phi_2)/2+\vartheta]}&
\nonumber\\
&\tilde{r}_{22}=-[\epsilon+(1-\epsilon)r]\,e^{i(\phi_2+\vartheta)}\,.&\nonumber
\end{eqnarray}
Out of this set of parameters, only two ($\epsilon,r$) 
are physically relevant.
The variable $\epsilon$ represents
the coupling of the incoming lead state to the edge state $n=1$
in the interior and $r\in {\mathbb R}$ is related 
to the reflection coefficient
$\tilde{r}''_{11}$ of an incident channel at the lead mouth, 
$\tilde{r}''_{11}=r e^{i(\psi+\vartheta)}$.  
Both quantities $\epsilon,|r|$ are restricted to the interval 
$[0,1]$. With these 
terms the absolute square of $t_{11}$ [Eq.~(\ref{3.5})] 
can be considerably simplified,\cite{kimfoot}
\begin{equation}\label{eq:fano}
T^{\rm tot}=|t_{11}|^2=\frac{(1-r^2)^2}{|\alpha|^2|\beta|^2}
\frac{\sin^2(\eta/2+\vartheta_0)}{\sin^2(\eta/2+\vartheta_0+\Delta)+\Gamma_0^2}\,,
\end{equation}
with $\phi=(\varphi_2+\phi_2)-(\varphi_1+\phi_1),\, 
\eta=(\varphi_2+\phi_2)+(\varphi_1+\phi_1),\,
r'=(1-\epsilon)\,e^{-i\phi/2}+\epsilon\,e^{i\phi/2},
\delta=\arg(r'),\,\vartheta_0=\vartheta+\delta,\,
\alpha=1+r\, e^{i(\eta+2\vartheta)},\,\beta=1+r e^{-2i\delta},\,
\Delta=\arg(\beta/\alpha)$. The linewidth $\Gamma_0$ 
is given by
\begin{equation}
\Gamma_0=\left|\frac{1-|r'\beta/\alpha|^2}{2r'\beta/\alpha}\right|\,.
\end{equation}
In the generic case of $r\neq 0$, resonances occuring in
Eq.~(\ref{eq:fano}) show a typical Fano profile of the form\cite{fano61} 
\begin{equation}
T^{\rm tot}\approx|t^{\rm bg}|^2\frac{(B/c-B_n/c)^2}
{(B/c-B_n/c+\Delta)^2+\Gamma_0^2}\,,
\end{equation} 
with $t^{\rm bg}$ being the coefficient for background scattering.\cite{nock94}
The Fano resonances at $B/c=B_n/c-\Delta$ 
will have an asymmetric lineshape
unless $\Delta=0$ (i.e.~$r=0$). This is however the case for the 
billiard systems we consider, since almost no reflection
of incoming lead states takes place at the lead mouths, 
$\tilde{r}''_{11}\approx 0$, and therefore $r\approx 0$.
Under this assumption Eq.~(\ref{eq:fano}) simplifies to
\begin{equation}\label{eq:fano2}
T^{\rm tot}\approx
\frac{\sin^2(\eta/2+\vartheta_0)}{\sin^2(\eta/2+\vartheta_0)+\Gamma_0^2}\,,
\end{equation}
with linewidth $\Gamma_0=(1-|r'|^2)/(2|r'|)$. This equation
describes symmetric
resonance lineshapes which can be identified as {\it window resonances} 
(also called Breit-Wigner dips/antiresonances) of the form
\begin{equation}
T^{\rm tot}\approx\frac{(B/c-B_n/c)^2}{(B/c-B_n/c)^2+\Gamma_0^2}\,.
\end{equation}
The physical picture resulting from this analysis is the following: In 
the magnetic field region $B^2_c \leq B \leq B^1_c$, where two 
edge states are present in the interior of the dot and one 
in each of the leads, the transmission probability 
shows large-scale oscillations 
intersected by sharp window resonances. The large-scale envelope
function is given by $1/(1+\Gamma_0^2)$. Its maxima
perfectly match with the roughly estimated term $\cos^2[B(A_1-A_2)/(2c)]$
from above and can therefore 
be identified with the numbered points in Fig.~\ref{fig:spectra}, 
each of which corresponds to an integer number of longitudinal 
antinodes in the wavefunction along the boundary (see
Fig.~\ref{fig:edgepics}). 
The antiresonances superimposed on these oscillations
occur at magnetic fields $B=B_n$ (where $\eta/2+\vartheta_0=n\pi,
\,n\in{\mathbb Z}$) and their linewidth is
given by $\Gamma_0$. As a result, resonances which
are situated on maxima of the term $1/(1+\Gamma_0^2)$ are sharper than
at its mimima [see numerical data in Fig.~\ref{fig:spectra} for confirmation].
For an increasing number of edge states populated in the cavity
($B<B_c^2$) 
our numerical results show that the density of antiresonances is
rapidly growing. This behaviour finally leads to a resonance overlap
for a large number of edge states, 
which is prerequisite for the
onset of Ericson fluctuations (i.e.~universal conductance 
fluctuations).\\For completeness we remark that the above analysis 
for the $B$-dependence of $T^{\rm tot}$ can likewise 
be carried out with $k_F$ instead of $B$ 
as the variable parameter. We can similarly identify threshold 
values $k_c^n$ for $k_F$, \textit{below} which a number 
of $n$ edge states survive.
The numerical results for the transmission probablility $T^{\rm tot}(k_F)$
for the case of one or two participating edge states (not shown)
can again be 
described by Eq.~(\ref{3.4}) or Eq.~(\ref{eq:fano}) respectively.
\subsection{Comparison with experiments}
A series of experiments\cite{marc92,wees89,webb85,pede00} have been 
performed where Aharonov-Bohm oscillations (ABOs) similar 
to the ones discussed here have been observed in ballistic 
transport measurements. The origin of the ABOs in these experiments
is however twofold: In Refs.~\onlinecite{marc92,wees89}, it is the presence
of edge states in a quantum dot which gives rise to the observed oscillations.
In Refs.~\onlinecite{webb85,pede00} on the contrary, the
investigated scattering devices have the form of a ring, to which the
scattering wave function is confined. The latter setup thus gives rise
to ABOs already at low magnetic fields and has therefore been 
more readily accessible to a
theoretical description.\cite{pich97} However, to our knowledge,
no quantitative description for magnetotransport through
a quantum dot in the regime of only one or two participating edge states 
has yet become available. 
We therefore discuss in the following 
similarities and differences between the experimental data and our 
calculations in this field. One important observation is that the 
magnetic fields where these quasi-regular transmission 
fluctuations appear in the experiment 
are lower than in the present calculation. 
For example, in the experiment for circle 
and stadium shaped quantum dots in a 
\textit{GaAs/AlGaAs}-heterostructure,\cite{marc92} 
the threshold magnetic field values 
would be (in SI-units) 
\begin{equation}
B_c^n=\frac{2\pi h}{(2n+1)\lambda_F^2e}\quad{\rm with}\quad 
\lambda_F=\sqrt{\frac{2\pi}{n_s}}\,.
\end{equation}
With a given sheet density of $n_s=3.6\times 10^{11} {\rm cm}^{-2}$
in the interior of the dot, the threshold magnetic fields are given by 
$B_c^2\approx 3$ Tesla and $B_c^1\approx 5$ Tesla. 
However, in the experiment regular oscillations
were already observed below
2 Tesla. At those field values we find highly 
irregular transmission fluctuations corresponding 
to a threshold magnetic field $B_c^n$ with $n \gg 1$, 
indicative of a high density of resonances and Ericson fluctuations. 
We expect the origin of this discrepancy to lie in the 
absence of sharp edges in the experiment and, hence, of diffractive edge 
scattering. In the experimental quantum dot, the edges 
should be fairly smooth, leading to near-adiabatic 
transitions to edge states at the entrance to the 
quantum dot. Therefore fewer edge channels are excited 
than by diffractive edge scattering, where all energetically 
accessible channels up to $n$ are populated. Our present 
results suggest that the observed transmission fluctuations 
are a direct measure of the sharpness of the edges at the 
lead mouth. Therefore, investigations of quantum dots with 
varying sharpness of edges would be desirable. Since these 
are, however, difficult
to fabricate  we point to a different 
experimental approach, which is based on the
analogy between transport in the edge state regime and field-free 
transport through a rectangle where
only few propagating modes participate. Such structures 
are accessible for microwave 
experiments.\cite{stoe99,blom02}
The measured transmission through such a microwave device
could provide a stringent test for the multi-channel interference
model presented above.
\section{SUMMARY AND OUTLOOK}
We have presented a new technique for calculating ballistic 
magnetotransport through
open quantum dots. The {\it Modular Recursive Green's Function
Method} (MRGM) is an extension of the widely used standard recursive 
Green's function technique and is based on the decomposition of 
non-separable scattering geometries
into separable substructures (modules). An unprecedented 
energy and magnetic field range can thereby be explored with high accuracy. 
We applied the MRGM to transport coefficients and 
scattering wavefunctions in the two extreme cases of high 
magnetic fields
and short wavelengths. For very small cyclotron radii we found 
periodic
oscillations in the transmission spectrum and beating phenomena,
which are restricted to well defined intervals
(as a function of $B$ and $k_F$ likewise). These features could be explained by
interferences between edge states, travelling along the 
boundary of the cavity.
For these states scattering only takes place at the lead 
junctions, whose
sharp edges play a crucial role for the dynamics of the system.
For a detailed analysis a multi-channel interference model 
was employed. This model allows to classify the observed 
transmission fluctuations in the framework of Fano resonances.
For only one edge state present in the circular dot 
transport is independent of 
the lead orientation provided that the
lead mouths are identical and separated from each other.  
Future envisioned applications include 
the investigation of Andreev billiards,\cite{cser02} 
quantum Hamiltonian ratchets,\cite{scha01}
fractal conductance fluctuations,\cite{ketz96,baec02}
and shot noise.\cite{ober02}  
Furthermore, the MRGM also seems suitable 
to perform ab-initio calculations of the integer 
Quantum Hall effect.\cite{kosc01} The challenge is in 
this case the inclusion of a disorder potential which is 
compatible with the separability conditions.
\begin{acknowledgements}
Helpful discussions with Prof.~Langer, C.~Stampfer and L.~Wirtz are
gratefully acknowledged. Many thanks are also due to W.~Gansterer 
and X.~Yang 
for their computer codes. This work was supported by the 
Austrian Science
Foundation (FWF).
\end{acknowledgements}

\newpage

\begin{figure}[!tbh] 
\caption{(color online) Absolute square of the
scattering wave functions $|\psi(x,y)|^2$ at high $k_F$ 
[(a): $k_F=25\pi/d$, (b): $k_F=12.5\pi/d$, (c),(d): $k_F=6.01\pi/d$] 
for the four quantum dots considered: circle and stadium 
with relative lead orientation of $90^\circ$ and $180^\circ$, area 
$A^{\rm dot}=4+\pi$ and lead width $d=0.25$. 
The localization around classical trajectories (see insets for comparison)
is clearly visible. In Figs. (a)-(c) the magnetic field 
$B=0$. In Fig. (d) the magnetic field $B/c=30.5$ 
allows for a whole bundle of equivalent trajectories
with cyclotron radius $r_c=k_Fc/B\approx 2.48$ to contribute
to transport.}
\label{fig:highenergy}
\end{figure}

\begin{figure}[!tbh] 
\caption{(color online) Absolute square of the
scattering wave functions $|\psi(x,y)|^2$ in the edge state
regime. 
The area of all geometries $A^{\rm dot}=4+\pi$, lead width $d=0.25$, and
$k_F=1.5\pi/d$. The four plots correspond to the points
in the transmission spectra (Fig.~\ref{fig:spectra}), indicated
by the letters (a)-(d). The numbers along the longitudinal direction 
of the edge states count the number of antinodes between entrance
and exit lead (see corresponding numbers in Fig.~\ref{fig:spectra}). 
Note that edge states in the magnetic field region $B_c^{n+1}<B<B_c^n$ have
up to $n$ transverse nodes: 
(a) circle, $180^\circ$, $n=2$, (b) 
circle, $90^\circ$, $n=2$, (c) stadium, $180^\circ$, $n=3$, and (d) stadium, 
$90^\circ$, $n=1$.}
\label{fig:edgepics}
\end{figure}

\begin{figure}[!tbh] 
\includegraphics[draft=false,keepaspectratio=true,clip,width=0.75\linewidth]{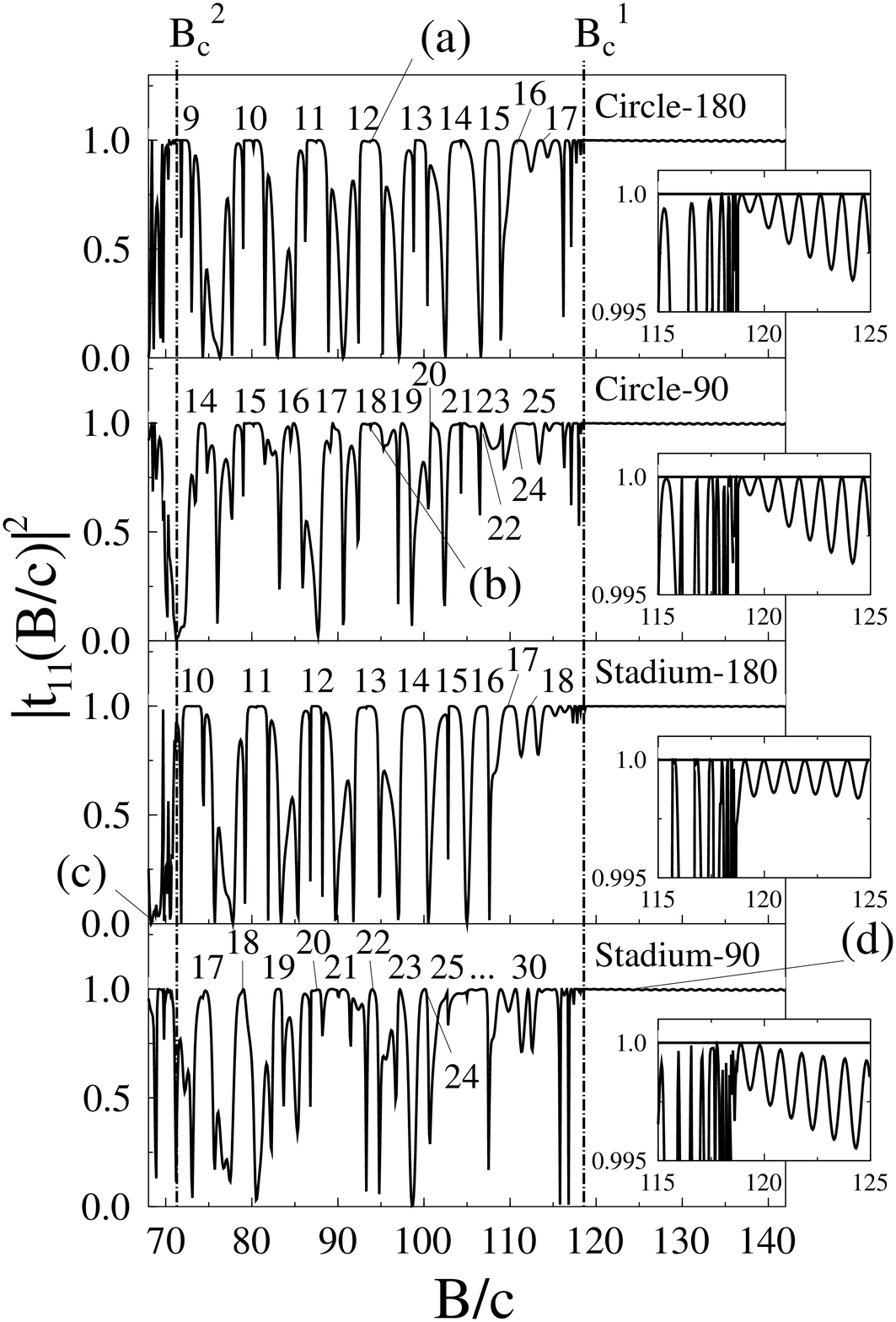}%
\caption{Transmission probabilities $|t_{11}(B/c)|^2$ in the
high magnetic field limit for circle and
stadium billiard with $180^\circ$ or $90^\circ$ lead orientation.
($k_F=1.5\pi/d,d=0.25$). $B_c^1$ and $B_c^2$ are the threshold magnetic
fields  $B_c^n/c=k_F^2/(2n+1)$ 
(vertical dash-dotted lines).
Above $B_c^1$ regular oscillations appear (see insets for magnification).
For $B_c^2<B<B_c^1$ irregular fluctuations set in.
Their large-scale
structure can be explained by the number of interference maxima
the two  edge states form along the boundary between 
entrance and exit lead (see indicated numbers). 
The points (a)-(d) correspond to the wavefunctions
shown in Fig.~\ref{fig:edgepics}.}
\label{fig:spectra}
\end{figure}

\begin{figure}[!tbh] 
\caption{(color online) 
Electron density $|\psi(x,y)|^2$ for the circle 
billiard with diffractive scattering highlighted.
($A^{\rm dot}=4+\pi$, lead width $d=0.25$ and
$k_F=1.5\pi/d$.) The magnetic field $B=118.7$ is just above the threshold 
to the single-edge state regime $B_c^1=118.44$.}
\label{fig:labelnew}
\end{figure}

\begin{thebibliography}{02}
\bibitem{kimfoot} The corresponding definitions in Ref.~\onlinecite{kim01}
contain a few errors, as was kindly pointed out to us by 
the principal author. These errors are corrected here.
\end{thebibliography}
\end{document}